\documentstyle[sprocl]{article}

\input{psfig}

\def\graphpath{}

\def\Journal#1#2#3#4{{#1} {\bf #2}, #3 (#4)}

\def\NPB{{\em Nucl. Phys.} B}
\def\PLB{{\em Phys. Lett.}  B}

\def\PRD{{\em Phys. Rev.} D}

\def\be{\begin{equation}}
\def\ee{\end{equation}}
\def\bea{\begin{eqnarray}}
\def\eea{\end{eqnarray}}

\def\d{\partial}
\def\uk{\underline{k}}

\begin{document}

\title{Testing SDLCQ in 2+1 dimensions}

\footnotetext{$^a$Based on Ref.~\cite{HHLPT99} in collaboration
with P.~Haney, J.R.~Hiller, O.~Lunin and S.S.~Pinsky.}

\author{U. TRITTMANN\footnotemark}

\address{Department of Physics\\Ohio State University\\174 W 18th Ave\\
Columbus, OH 43210-1106, USA}


\maketitle\abstracts{ 
The SDLCQ regularization is known to 
explicitly preserve supersymmetry in 1+1 dimensions.
To test this property in higher dimensions,
we consider supersymmetric Yang-Mills theory on $R\times S^1\times S^1$. 
In particular, we choose one of the compact directions to be light-like 
and another to be space-like. 
This theory is totally finite, and thus 
we can solve for bound state wave functions and masses numerically without 
renormalizing. We present the masses as functions of
the longitudinal and transverse resolutions and show that the masses 
converge rapidly in both resolutions. We study the behavior of the 
spectrum as a function of the coupling and find that at strong coupling 
there is a stable, well-defined spectrum which we present. 
We discuss also the massless spectrum and find several
unphysical states that decouple at large transverse resolution. 
}

\footnotetext{Supported by an Ohio State University Postdoctoral Fellowship.}


\section{Introduction}

The motivations to 
consider ${\cal N}=1$ supersymmetric Yang-Mills theories in 2+1 dimensions 
are manifold.
For one, there is recent progress in understanding the properties of strongly 
coupled gauge theories with supersymmetry
\cite{seibergwitten}, some of which are believed
to be interconnected through a web of strong-weak coupling dualities.
There is a need to study these issues at a fundamental level. Ideally,
we would like to solve for the bound states of these
theories directly, and at any coupling.
However, solving a field theory from first principles
is typically an intractable task. Nevertheless, it has been
known for some time that $1+1$ dimensional field theories
can be solved from first principles via a straightforward
application of DLCQ \cite{bpp98}.
Recently, a large class of supersymmetric
gauge theories in two dimensions was studied using
a supersymmetric form of DLCQ (SDLCQ), which is known to
preserve supersymmetry at every stage of the calculation
\cite{sakai95,alp98a}.
It turns out that this formalism can be applied
to higher-dimensional theories \cite{alp99b}.
This is interesting, because in higher dimensions, due 
to the additional scale,
theories have the potential of exhibiting a complex phase structure,
which may include a strong-weak coupling duality.

\section{SDLCQ}

We consider a three dimensional SU($N_c$) ${\cal N}{=}1$
super-Yang-Mills theory 
compactified on the space-time ${\bf R} \times S^1 \times
S^1$. The calculations are performed in the large $N_c$ limit.  In
particular, we use light-cone coordinates 
$x^{\pm}=\frac{1}{\sqrt{2}}(x^0\pm x^1)$, 
compactify $x^-$ on a light-like circle {\em a la} DLCQ, 
and wrap the remaining  transverse
coordinate $x^{\perp}$ on a spatial circle. We are able to solve
for bound state wave functions and masses numerically by diagonalizing the 
discretized light-cone supercharge. 
This procedure preserves supersymmetry at every step. 
The action of  ${\cal N}=1$ SYM(2+1) is
\[
S=\int d^2 x \int_0^L dx_\perp \mbox{tr}(-\frac{1}{4}F^{\mu\nu}F_{\mu\nu}+
{\rm i}{\bar\Psi}\gamma^\mu D_\mu\Psi).
\]
We decompose the spinor $\Psi$ in terms of chiral projections $\psi, \chi$
and choose the light-cone gauge $A^+=0$.
We solve for the non-dynamical fields $A^-$ and $\chi$
and formulate the momentum operators in the physical degrees of freedom 
($\phi\equiv A^2$)
\begin{eqnarray*}
P^+&=&\int dx^-\int_0^L dx_\perp\mbox{tr}\left[(\d_-\phi)^2+
{\rm i}\psi\d_-\psi\right],\\
P^-&=&\int dx^-\int_0^L dx_\perp\mbox{tr}
\left[-\frac{1}{2}J\frac{1}{\d_-^2}J-
            \frac{{\rm i}}{2}D_\perp\psi\frac{1}{\d_-}D_\perp\psi\right].
\end{eqnarray*}
The canonical commutation relations yield the supersymmetry algebra
\[
\{Q^+,Q^+\}=2\sqrt{2}P^+, \{Q^-,Q^-\}=2\sqrt{2}P^-,
\{Q^+,Q^-\}=-4P_\perp.
\]
We use the standard decomposition of the fields $\phi_{ij}(x^-,x^{\perp})$ and 
$\psi_{ij}(x^-,x^{\perp})$
into momentum modes $a_{ij}^{\dagger}(\underline{k})$ and 
$b_{ij}^{\dagger}(\underline{k})$, respectively. We defined 
$\uk\equiv(k^+,n^{\perp})$ for convenience. 
The (anti-)commutation relations 
\begin{eqnarray*}
&&\left[a_{ij}(\uk),a^\dagger_{lk}(\uk')\right]=
\left\{b_{ij}(\uk),b^\dagger_{lk}(\uk')\right\}=
\delta(k^+ -k^{'+})\delta_{n^\perp,n^{'\perp}}\delta_{il}\delta_{jk}.
\end{eqnarray*}
yield the explicit form of the supercharges, which are listed in 
Ref.~\cite{HHLPT99}. 
For the present discussion it suffices to know that the structure 
of $Q^-$ is
%
\begin{eqnarray*}
\label{Qminus}
Q^-&=&\frac{2^{3/4}\pi {\rm i}}{L}\sum_{n^{\perp}\in {\bf Z}}\int_0^\infty dk^+
\frac{n^{\perp}}{\sqrt{k^+}}\left[
a_{ij}^\dagger(\uk) b_{ij}(\uk)-
b_{ij}^\dagger(\uk) a_{ij}(\uk)\right]+g \tilde{Q}^-,
%
%
%
\end{eqnarray*}
where $\tilde{Q}^-$ contains the terms with three operators.
We note also that the supercharge $Q^-$ is linear in the coupling $g$, and thus
the Hamiltonian $P^-$ is quadratic in $g$.

\begin{figure}\label{Figure1}
\centerline{
\psfig{file=\graphpath 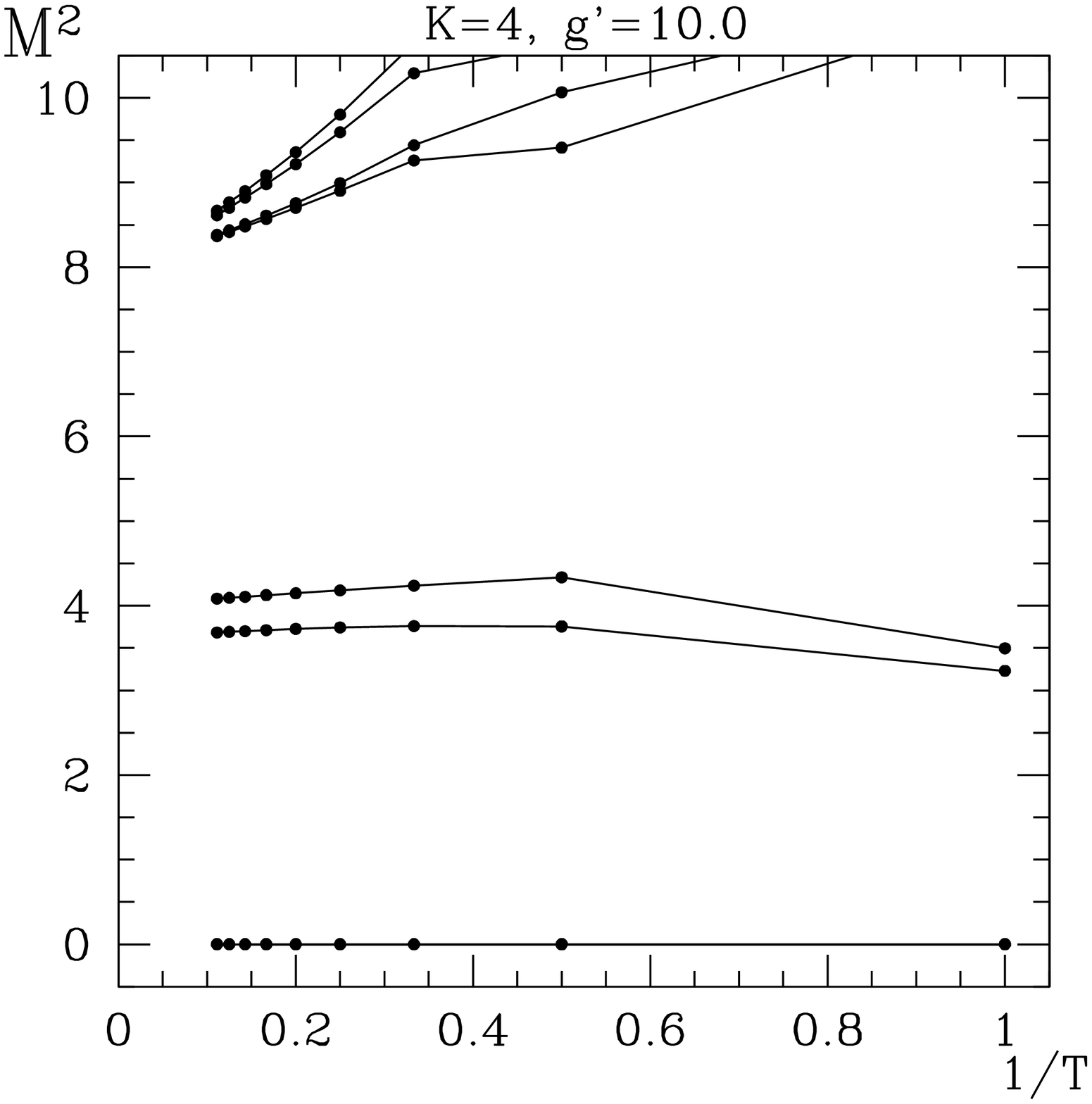,width=6true cm}
\psfig{file=\graphpath 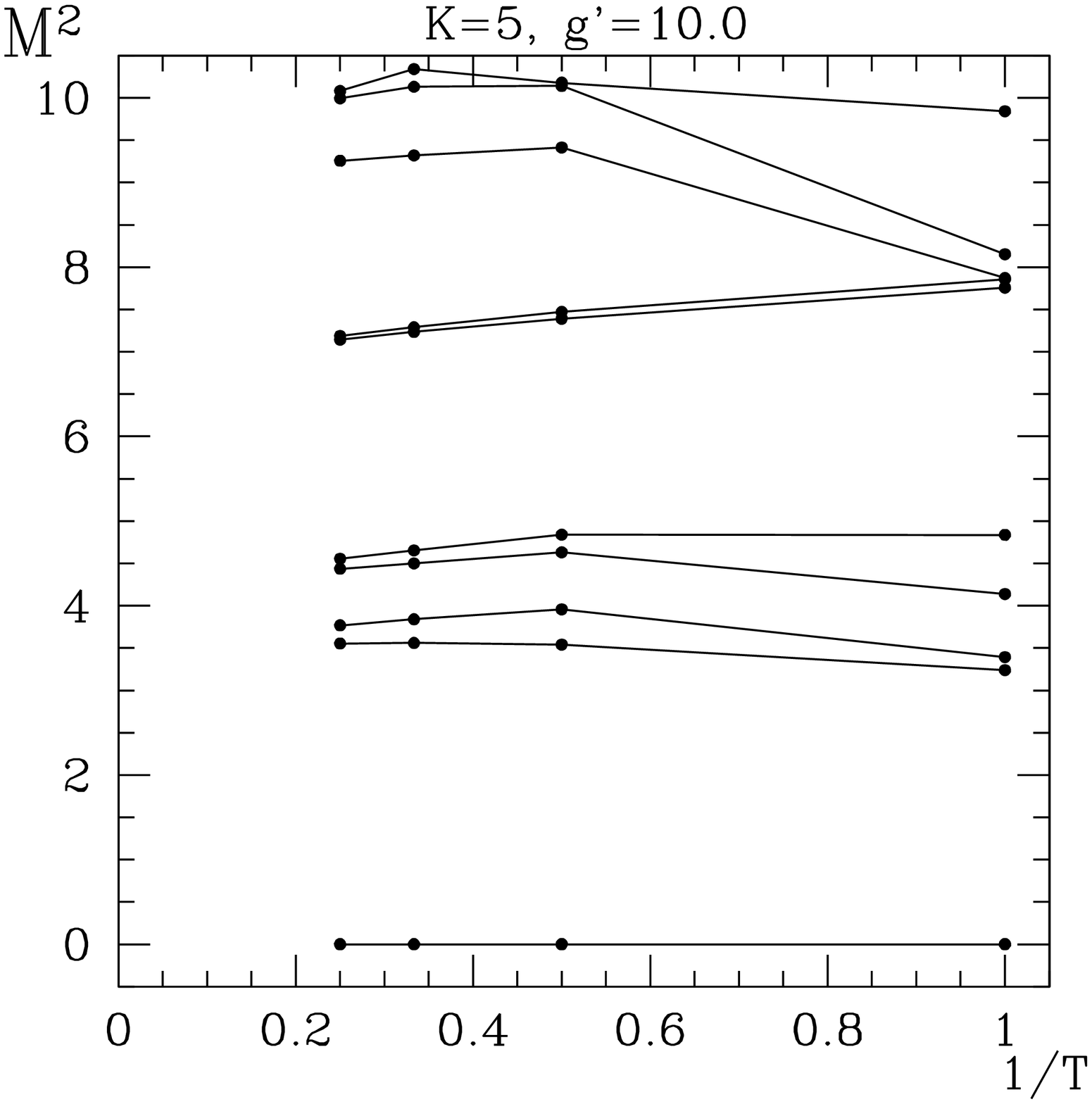,width=6true cm}
}
\caption{{\small Plot of bound state mass squared $M^2$
in units of $2 \pi^2/ L^2$  as a function of the transverse
resolution $T$ for a coupling
$g'=10$ and for longitudinal resolutions $K=4$ (a) and $K=5$ (b).
Boson and fermion masses are identical. }
}
\end{figure}
\vspace*{-0.5cm}

We now perform the truncation procedure.
The harmonic resolution $K$ plays the role of a longitudinal cutoff as usual,
and longitudinal momentum fractions take values
$\frac{k^+_i}{P^+}=\frac{n_i}{K}, n_i=1,2,\ldots,K$.
The transverse cutoff $T$ allows for 
momenta $k_i^{\perp}=2\pi n_i^{\perp}/L$
with $n_i^{\perp}=0,\pm 1,\pm 2,\ldots \pm T$. 
This prescription preserves parity symmetry in transverse directions.
How does such a truncation affect the supersymmetry properties of the
theory? Note first that an operator relation $[A,B]=C$ in the  full
theory is not
expected to hold in the truncated formulation.  However, if A is quadratic in
terms of fields (or in terms of creation and  annihilation operators), one can
show that the relation $[A,B]=C$ implies
$[A_{tr},B_{tr}]=C_{tr}$
for the truncated operators $A_{tr}$,$B_{tr}$, and $C_{tr}$.  In our
case, $Q^+$
is quadratic, and so the relations
$\{Q_{tr}^+,Q_{tr}^+\}=2\sqrt{2}P_{tr}^+$ and
$\{Q_{tr}^+,Q_{tr}^-\}=0$ are true
in the $P_\perp=0$ sector of the truncated theory.  The anticommutator
$\{Q_{tr}^-,Q_{tr}^-\}$,
however, is not equal to $2\sqrt{2}P_{tr}^-$. So the diagonalization of
$\{Q_{tr}^-,Q_{tr}^-\}$ will yield a different bound-state spectrum
than the one
obtained after diagonalizing $2\sqrt{2}P_{tr}^-$. Of course, the two spectra
should agree in the limit
$T\rightarrow\infty$. At any finite truncation, however, we have the liberty to
diagonalize either of these operators. The choice of $\{Q_{tr}^-,Q_{tr}^-\}$
specifies our regularization scheme.
Choosing to diagonalize the light-cone
supercharge $Q_{tr}^-$ has an important advantage:
{\em the spectrum is exactly supersymmetric for
any truncation}. In contrast, the spectrum of the Hamiltonian $P_{tr}^-$
becomes supersymmetric only in the infinite resolution limit.

Let us take a look at the discrete symmetries of $Q^-$.
The three commuting symmetries $Z_2$ are parity in the transverse direction
\[
P: a_{ij}(k,n^\perp)\rightarrow -a_{ij}(k,-n^\perp),\qquad
        b_{ij}(k,n^\perp)\rightarrow b_{ij}(k,-n^\perp),
\]
which anti-commutes with $Q^+$ and $P_\perp$, and
a generalized $T$ symmetry
\[
S: a_{ij}(k,n^\perp)\rightarrow -a_{ji}(k,n^\perp),\qquad
        b_{ij}(k,n^\perp)\rightarrow -b_{ji}(k,-n^\perp).
\]
To close the group, we need a third symmetry, namely $R=PS$.
An interesting detail of the symmetry considerations is the fact that the
$P$ symmetry leads to exactly degenerate eigenvalues. This means in turn that 
all massive eigenvalues are four-fold degenerate.
The argument goes as follows.
Start with a massive state with positive parity $|M+\rangle$ which obeys
\[
(Q^-)^2|M+\rangle=M^2|M+\rangle, \qquad
P|M+\rangle=+|M+\rangle.
\]
Then $Q^+Q^-|M+\rangle$ is a state with same mass but opposite parity
\[
PQ^+Q^-|M+\rangle=-Q^+Q^-P|M+\rangle=-Q^+Q^-|M+\rangle.
\]

\begin{figure}\label{Fig2}
\centerline{
\psfig{file=\graphpath 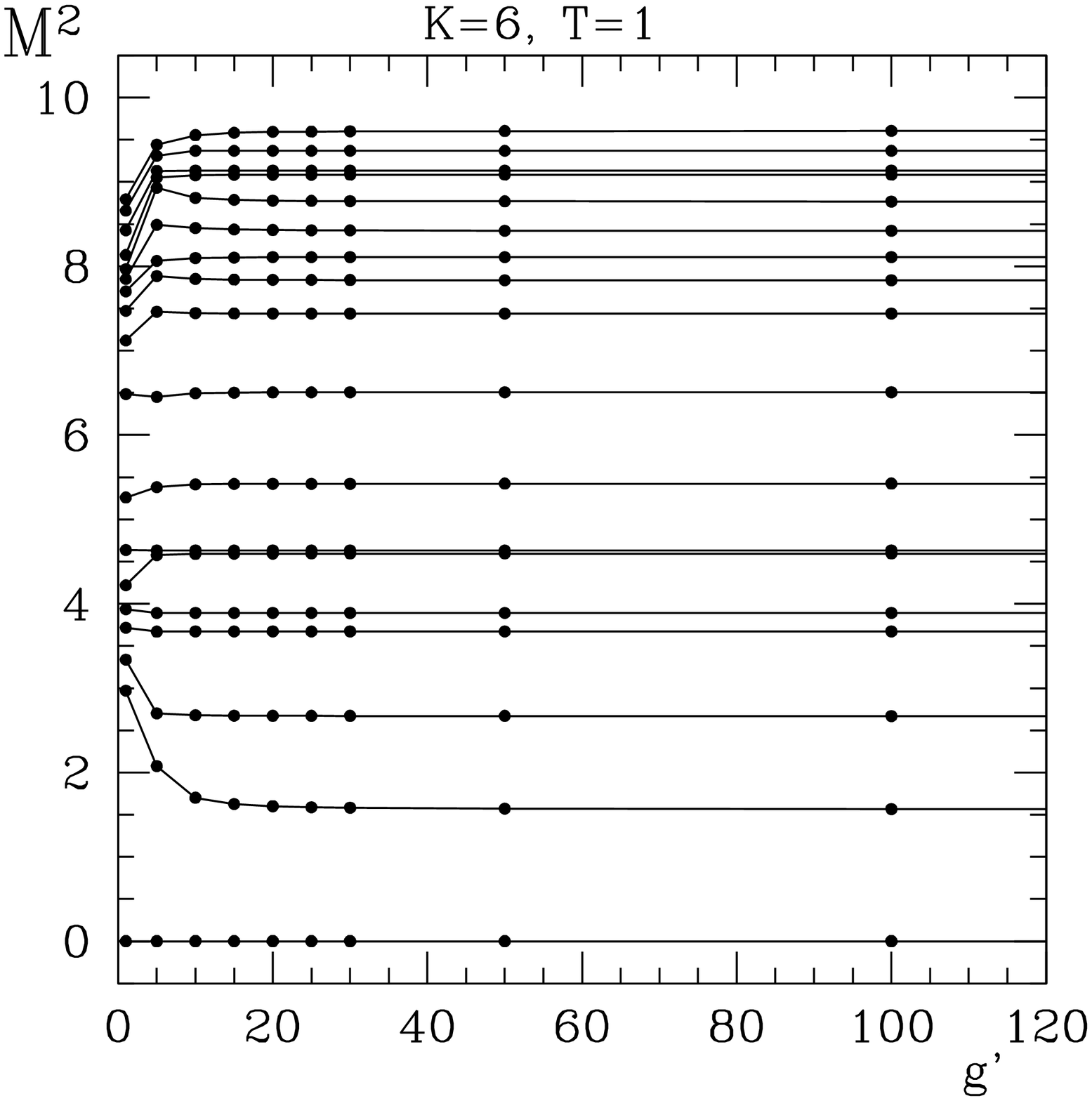,width=6true cm}
\psfig{file=\graphpath 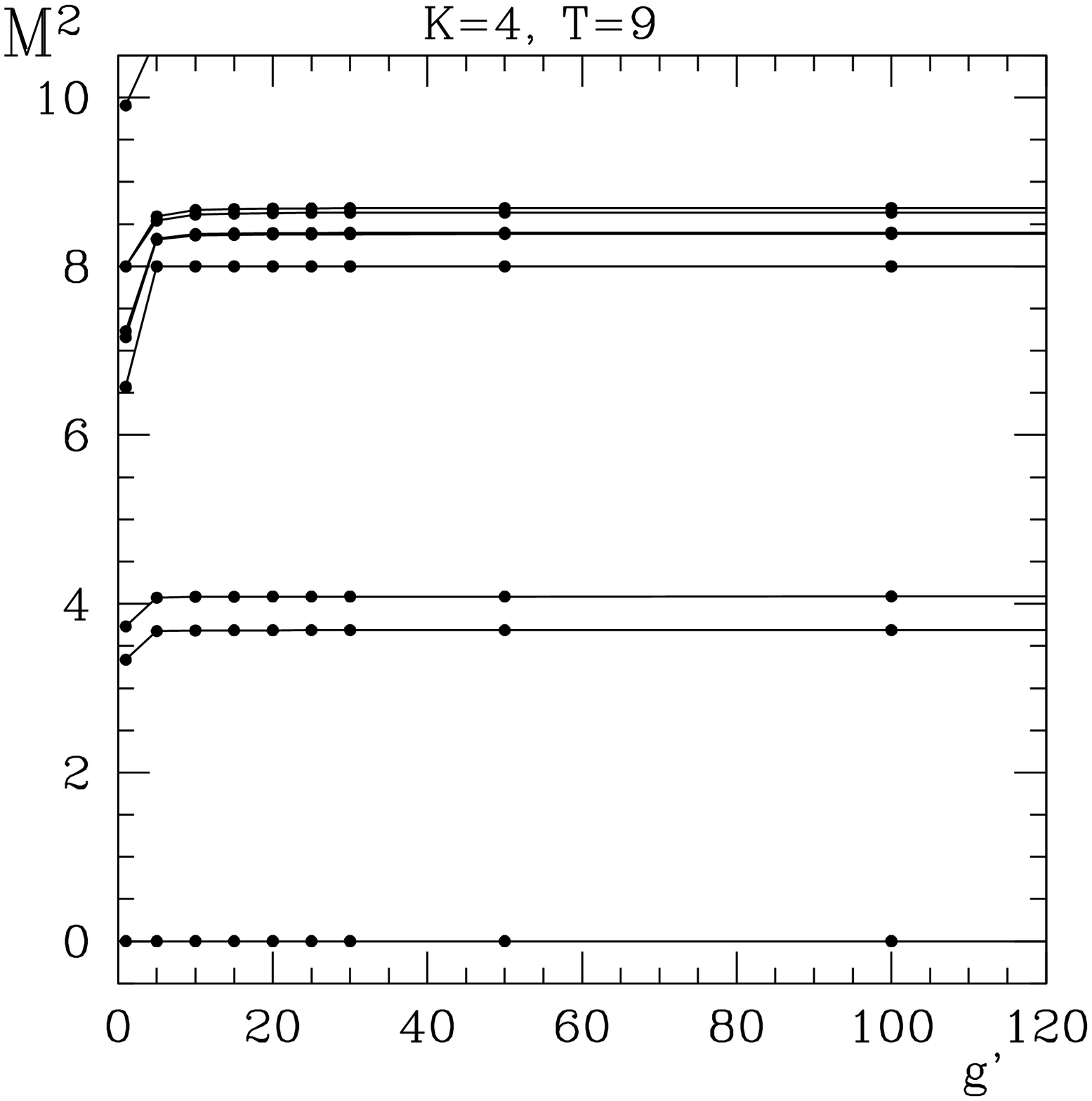,width=6true cm}}
\caption{{\small  Plot of bound state mass squared $M^2$
in units of $2 \pi^2 / L^2$  as a function of the coupling
$g'$.  We show the plots for $K=4$, $T=9$ (a) and $K=6$,
$T=1$ (b).}}
\end{figure}
\vspace*{-0.5cm}

\section{Numerical Results}

With the truncation prescription described above, we can solve the 
discretized eigenvalue problem 
\[
2P^+P^-|\psi\rangle=M^2|\psi\rangle,
\]
characterized by the cutoffs $K$ and $T$, on the computer. 
This is equivalent to constructing
the supercharge $Q^-$ in the usual Fock basis, and then diagonalizing it.
If the resulting mass (squared) eigenvalues $M^2$ are plotted as a function 
of the dimensionless coupling $g'=g\sqrt{NL/4\pi^3}$, several striking 
features emerge. Namely, as was noted in
Ref.~\cite{alp99b}, one finds a very stable 
strong-coupling spectrum. Secondly, we find states which fall off 
fast to zero mass with increasing coupling.
Since the previous work \cite{alp99b} was a calculation of the spectrum 
at $T\equiv 1$, it is natural to ask whether
the well defined large $g'$ spectrum survives at $T\rightarrow\infty$,
and to study the properties of the states with masses decreasing at large 
$g'$. A further question is if the number of massless states is independent 
of the transverse cutoff $T$.

Our previous SDLCQ calculations were done using a code written in
Mathematica and performed on a PC. This code has now been rewritten in
C++ and some of the present work was done on supercomputers. We were able
to perform numerical diagonalizations for $K=2$ through 7 and for values of
$T$ up to $T=9$ at $K=4$ and $T=1$ at $K=7$.

\underline{\bf Massive spectrum:}
Little is known about the large coupling spectrum of quantum field theories, 
with the exception of theories in 1+1 dimensions. There, however, the
concept of large coupling has no meaning, since the coupling is only a 
multiplicative constant in the Hamiltonian.
In particular, there is no weak/strong duality known in 
${\cal N}=1$ SYM(2+1), which could give some clue how the spectrum looks
like.
We performed therefore a non-perturbative calculation in SDLCQ
to directly access the spectrum.

In Fig.~\ref{Figure1} we plot the bound state masses squared $M^2$ as 
a function of the transverse resolution $T$ for $K=4$ and $K=5$. 
We see that the curves are very flat, thus exhibiting fast convergence 
in transverse cutoff $T$.
The continuum result can be obtained by extrapolating the curves to 
$1/T\rightarrow0$.
Let us look at the bound states as a function of the coupling $g'$,
focusing on the large coupling regime.
We see, Fig.~\ref{Fig2}, that the states are extremely stable in $g'$, 
{\em i.e.} they are quasi independent of coupling.
We find this behavior at every value of $K$, and even irrespective of the 
value of the transverse cutoff $T$.
\begin{figure}
\centerline{
\psfig{file=\graphpath 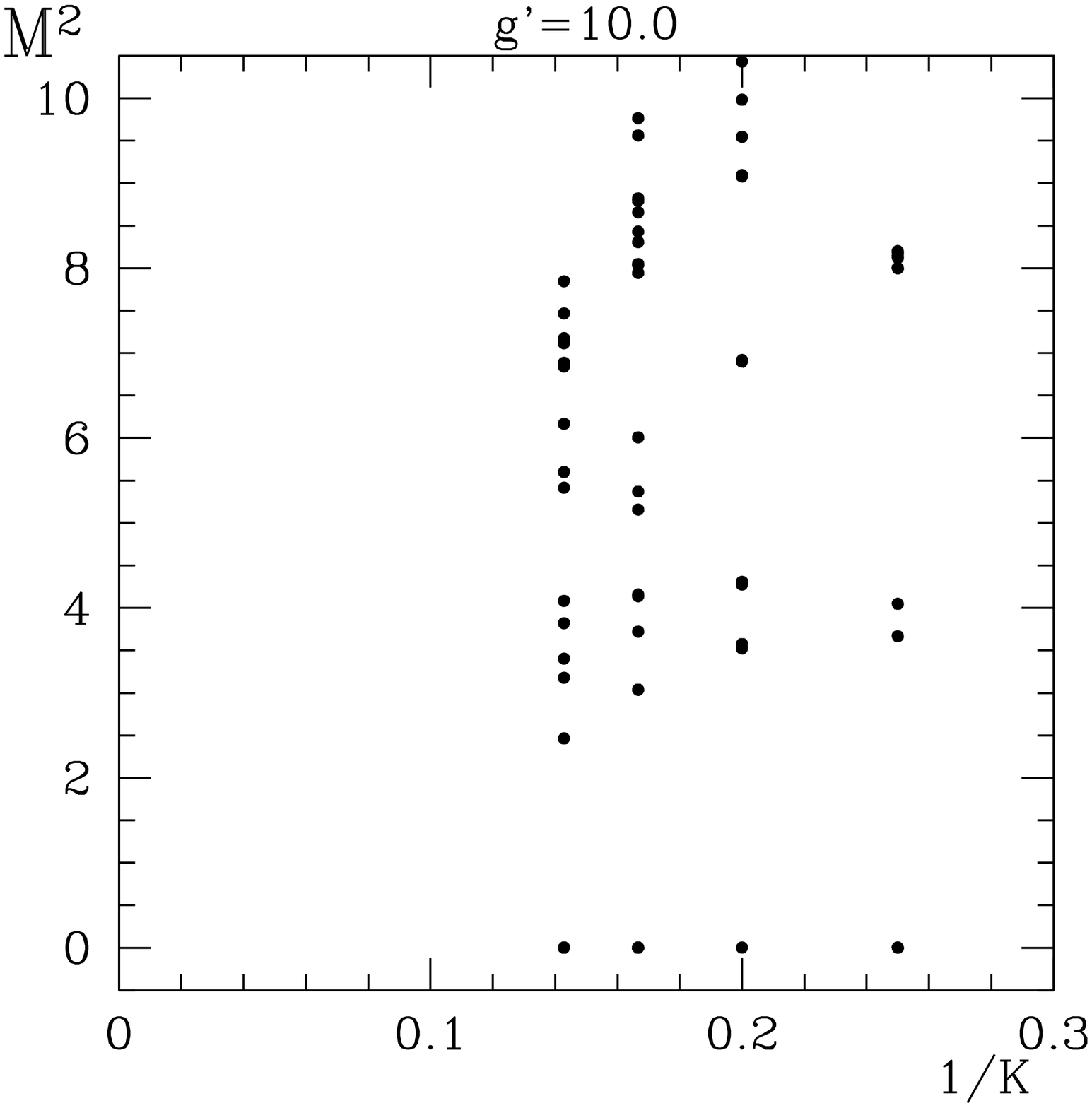,width=6true cm}
\psfig{file=\graphpath 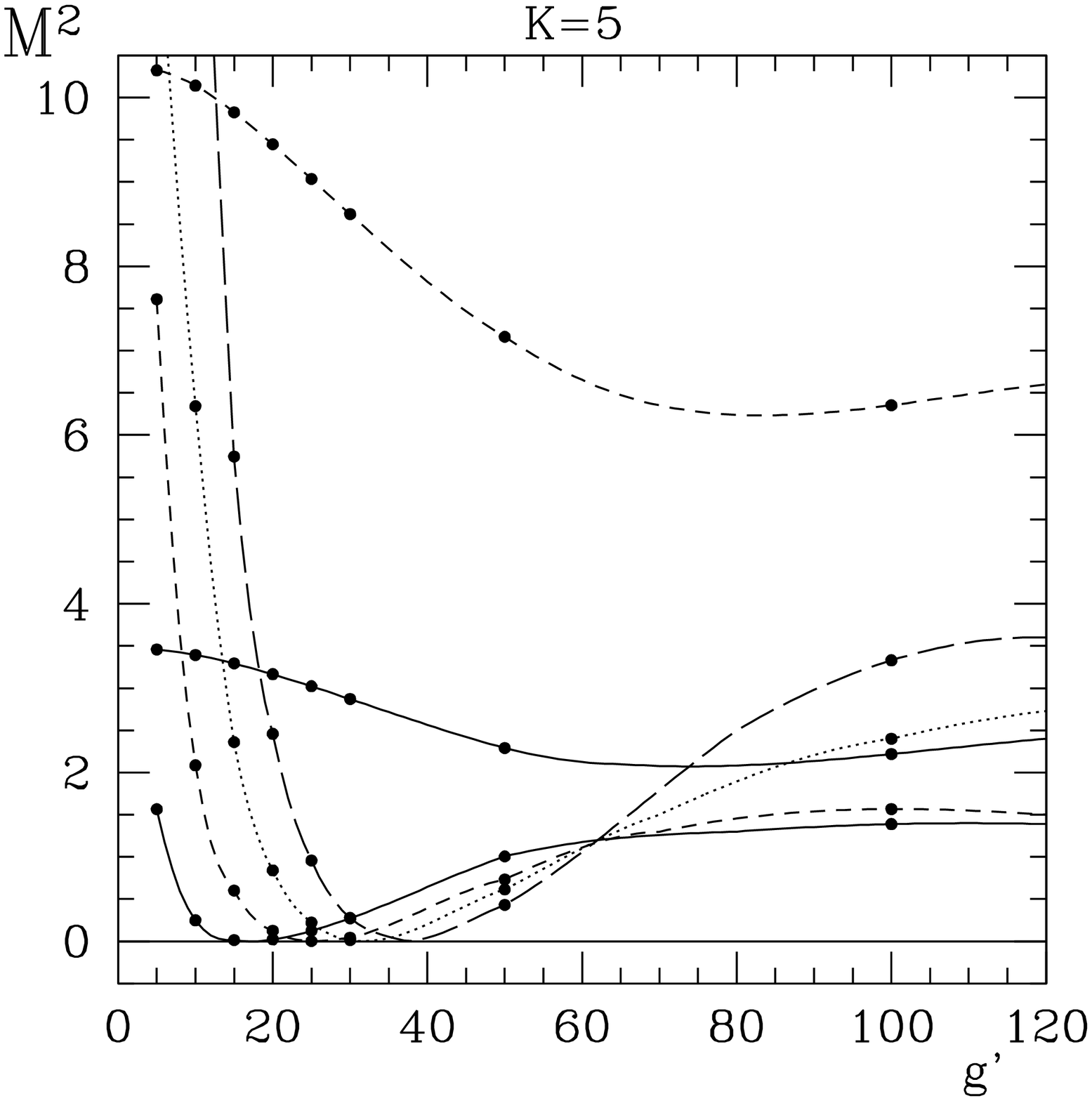,width=6true cm}
}
\caption{{\small Plot of bound state mass squared $M^2$
in units of $2 \pi^2 / L^2$  as a function of $1/K$ for coupling
$g'=10$ (a).  
States with masses falling rapidly with
increasing coupling as a function of $g'$ (b)
at the transverse resolutions $T=1$ (solid lines),
$T=2$ (dashed lines), $T=3$ (short dashed line) and $T=4$ (long dashed line).
}}
\label{Fig3}
\end{figure}
We show the bound state mass as a function of $1/K$ in Fig.~\ref{Fig3}(a).
These results are the first calculation of the
strong-coupling  bound states of ${\cal N}=1$
SYM in $2+1$ dimensions. As we increase
the resolution we are able to see states that have, 
as their primary component,
more and more partons, and, as we have seen in other supersymmetric
theories, many
of these states appear at low energies.  This accumulation of
high-multiplicity
low-mass states
appears to be a unique property of SUSY theories. In non-SUSY theories the
new states appear at increasing  energies. In the dimensionally reduced
version of this theory we saw that the
accumulation point of these low-mass states
appeared to be at zero mass
\cite{alp99b,ahlp99}. Here again we see clear evidence of an
accumulation of low
mass states, however we don't have sufficient information to say whether an
accumulation point exists.

\underline{\bf Massless states:} 
There are two kinds of massless states in the spectrum.
Firstly, the states massless for $g'\rightarrow 0$. They are 
massless because at small coupling, only first term of the supercharge, 
Eq.~(\ref{Qminus}), gives a contribution. Then all partons 
with $n^{\perp}=0$ (anti-)commute with $Q^-$. Thus all states made out
of these partons are massless, {\em and} massless states contain 
just these partons. 
The set of massless states at $g'=0$ therefore 
coincides with a Hilbert space of the theory dimensionally reduced to $1+1$.
Moreover, the whole infrared spectrum of SYM${2+1}$ at small coupling is
governed by the dimensionally reduced theory (see \cite{alp99b} for details)
Secondly, we see states that are exactly massless at any coupling.
These are $2(K{-}1)$ BPS states, fulfilling 
$Q^-|m{=}0\rangle=0$, $Q^+|m{=}0\rangle\neq 0.$
It is therefore easy to construct the massless states of (2+1) theories, 
at least at large $N_c$.

\underline{\bf Unphysical states:}
Finally, we can unambiguously detect unphysical states due to their 
special properties.
Namely, these states vary strongly with coupling $g'$ and 
they appear (predominantly) for $K$ odd and 
decouple for $T\rightarrow \infty$, see Fig.~\ref{Fig3}.
It is thus easy to classify and to remove the unphysical states 
from the spectrum.


\section{Conclusions}

We have shown that the SDLCQ formalism naturally extends to higher dimensions.
Rapid convergence in transverse direction is found.
Concerning the specific theory, we obtained the strong coupling spectrum of 
${\cal N}=1$ SYM(2+1). The bound states are extremely stable for 
$g'\rightarrow \infty$.
The unphysical states in the spectrum can be identified and removed.
We found no new massless states at strong coupling compared to previous work 
\cite{alp99b}.
The massless sector of the theory is completely determined by the dimensional 
reduced model.
The light states turn out to be string-like and might contain
physics of dual theories.
Also, the theory might be conformal at decompactification limit.
We are currently working on analytical and numerical improvements
of the approach. We expect to be able to address problems 
like the very interesting ${\cal N}=4$ SYM in 3+1 and 
${\cal N}=1$ SYM(2+1) including a Chern-Simons term conjectured to
break supersymmetry, in the near future.

\section*{Acknowledgments}
It is a pleasure to thank the workshop organizers
for hospitality and support.

\end{document}